# Ultra-resolution photochemical sensing

Lukas Fürst (1), Alexander Eber (1), Mithun Pal (1), Emily Hruska (1), Clemens Hofmann (2), Iouli Gordon (3), Martin Schultze (1), Rolf Breinbauer (2), Birgitta Bernhardt (1)*

*Correspondence to: bernhardt@tugraz.at

(1) Institute of Experimental Physics, Graz University of Technology, Graz, Austria

(2) Institute of Organic Chemistry, Graz University of Technology, Graz, Austria

(3) Harvard-Smithsonian Center for Astrophysics, Atomic and Molecular Physics, Cambridge, USA

## Abstract

Photochemistry in the earth's atmosphere is driven by the sun, continuously altering the concentration and spatial distribution of pollutants. Precisely monitoring their atmospheric abundance relies predominantly on optical sensing, which requires the knowledge of exact absorption cross sections. One key pollutant which impacts many photochemical reaction-pathways is formaldehyde. Agreement on formaldehyde absolute absorption cross section remains elusive in the photochemically-relevant ultraviolet spectral region, hampering sensitive concentration tracking. Here, we introduce free-running ultraviolet dual comb spectroscopy, combining high spectral resolution (1 GHz), broad spectral coverage (12 THz), and fast acquisition speed (500 ms), as a novel method for absolute absorption cross section determination with unprecedented fidelity. Within this bandwidth, our method uncovers almost one order of magnitude more rovibrational transitions than detected before which leads to refined rotational constants for high-level quantum simulations of molecular eigenstates. This ultra-resolution method can be generalized to provide a universal tool for fast electronic fingerprinting of atmospherically-relevant species, both for sensing applications and to benchmark improvements of *ab-initio* quantum theory.

## 1. Introduction

Air pollution stands as one of the foremost environmental challenges of the 21st century, contributing to over 5 million deaths each year by adversely affecting human health, ecosystems, and climate change (*1, 2*). Despite regulations that have resulted in substantial reduction of pollution in cities in the last decade, the health hazards remain. In response to the widespread and immediate consequences of air pollution, the European Council updated its ambient air quality directives in October 2024 to enforce stricter standards (*3*). However, to effectively comply with the new regulations, it is essential to enhance our understanding of the earth's atmospheric condition. Both natural and anthropogenic activities influence its complex molecular composition, and with the sun driving photochemical reaction chains on multiple time-scales, it becomes necessary to monitor the fluctuating dynamics between interacting pollutant species.

Among all air pollutants, formaldehyde (HCHO) is a key volatile organic compound (VOC) that plays a pivotal role in atmospheric photochemistry since it serves as a precursor to both hydroxyl radicals ($HO_x$) and nitrogen oxides ($NO_x$) and reacts to form secondary organic aerosols and tropospheric ozone ($O_3$) (*4, 5*). HCHO also holds a central role in the prebiotic chemistry of life's key building blocks in the form of amino acids, nucleobases, and sugars; this motivates formaldehyde tracing in interstellar space (*6–8*). Naturally elevated concentrations of HCHO are frequently detected in forested areas and regions susceptible to biomass burning, where it serves as an indicator of organic combustion processes (*9*). However, anthropogenic activities, such as industrial emissions and vehicle exhaust, also act as sources for increased concentrations. Due to its toxicity and potential carcinogenic effects (*10, 11*), the air quality guidelines of the World Health Organization (WHO) establish a concentration limit of 0.1 mg/m³ to mitigate sensory irritation risks (*12*). Furthermore, HCHO poses threats to ecosystems and biodiversity by negatively impacting plant health, diminishing crop yields, and harming aquatic organisms through both atmospheric deposition and industrial emission (*13*). It is not a surprise that formaldehyde is one of the priority targets for multiple satellite and ground-based spectrometers operating in the ultraviolet (UV) (*14–17*). This multifaceted impact underscores the importance of understanding, locating, and tracking formaldehyde emissions for public health, environmental management, and fundamental research.

In atmospheric chemistry, highly energetic UV light serves as a catalyst for an extensive array of photochemical reactions involving HCHO and other gaseous species (e.g., $O_3$, $NO_2$, and NO). In tropospheric ozone photochemistry, the reaction cascade following $O_3$ photodissociation evolves via hydrogen radicals and HCHO, and their interaction with UV solar wavelengths between 340 nm and 365 nm constitutes a turning point of the reaction. When HCHO absorbs radiation below 365 nm, the cycle can terminate with a reduction of hydrogen radicals in ~ 2/3 of the reaction processes. Absorption of shorter wavelength radiation, however, evokes the production of methane and ammonia with a contrary effect of increasing the number of harmful hydrogen radicals $HO_x$ (~1/3 of the processes) (*18*).

Consequently, it is important to have reliable reference spectroscopic data for formaldehyde in this particular spectral region as they are one of the key input parameters for various atmospheric simulations (*19*) and fitting routines for satellite-based observations (*14, 15*). The HITRAN molecular spectroscopic database (*20*) provides line-by-line parameterization of formaldehyde only in the microwave and infrared spectral region, whereas for the UV part of the spectrum, only experimental cross-section can be used due to the difficulty in building a consistent and complete quantum mechanical model. Despite their importance, experimentally determined absorption cross sections known to date differ up to 20 %, highlighting the challenge for absolute accuracy (*21*). Probable origins of this uncertainty include the complexity of the potential reaction pathways with several branching reaction products, the tendency of formaldehyde to polymerize, and its affinity to surface adsorption. For the HITRAN database, Chance and Orphal (*22*) have recommended the use of cross-sections derived from the work of (*23*) with modifications (which included scaling of the intensities and introducing an offset) to achieve a better agreement with the lower resolution cross-sections from (*24*). To address challenges in obtaining reliable data experimentally, in addition to high resolution and sensitivity, extremely careful sample preparation and ultrashort acquisition times to minimize concentration changes during the measurement are requirements.

In this work, we introduce ultra-resolution UV dual comb spectroscopy (UV DCS), uniting all four of these prerequisites for the first time by employing passively-stable laser frequency combs without the need for complex cascaded frequency stabilization schemes. Recently, dual comb spectroscopy, combining elevated spectral resolution with short measurement times, has been extended to the UV spectral region (*25–29*), opening new spectroscopic capabilities for the investigation of byzantine molecular spectra. All implementations in the UV require the long-term coherent superposition of two optical frequency combs. Due to the high frequency of UV radiation and the strong dispersion and absorption of all optical elements in this wavelength range, achieving mutual coherence requires extensive active stabilization schemes or sophisticated feed-forward techniques (30–32). Here, we have developed an ultraviolet free-running dual-comb spectrometer for the first time without the need for active feedback, allowing the measurement of absorption spectra with 1 GHz frequency comb resolution in less than a second acquisition time.

The detection scheme achieves a factor of 20 improvement in the noise level, enhanced spectral resolution and sensitivity compared to the best available data known to us (18). Resulting from this advance, our method resolves almost one order of magnitude more rovibrational transition energies than have so far been experimentally observed (22). These rovibronic transitions have been theoretically predicted but, so far, evaded conclusive experimental observation due to their narrow linewidths and low transition strength.

To provide theory-driven studies with absolute, rather than relative, transition probabilities for the refinement of empirical simulations and the understanding of atmospheric photochemistry, extremely particular sample conditions must be met. First, the knowledge of the exact number density in the measured sample is required. Additionally, the presence of other molecular contaminants, which may ultimately contribute their own spectral features to the absorption signal or offer novel reaction pathways, must be minimized. Here, we ensure these specifications are met by preparing pure HCHO in the gas phase via an innovative double-distillation synthesis scheme.

### 2. Experimental methods

Figure 1 depicts the experimental scheme of free-running UV DCS. A Yb:CALGO crystal-based, single-cavity dual oscillator serves as the dual-comb source. The free-running laser system operates without active stabilization of the comb parameters, i.e., the repetition rate and carrier-envelope-offset frequency. Two optical frequency combs from a single laser cavity are spatially multiplexed using a Brewster-angled biprism (33). Consequently, both laser pulse trains have slightly different repetition rates. The third harmonic of the fundamental infrared light is generated using sum frequency generation in nonlinear crystals (see Methods). Both ultraviolet beams, centered at 855.4 THz (350.5 nm wavelength), are combined on a plate beamsplitter. The overlapped beams pass through the HCHO sample cell (34–36).

For the determination of the absolute absorption cross section, HCHO is prepared in an advanced two-step purification process to provide precise knowledge about the sample concentration with calibrated pressure gauges (see Supplementary). The time-domain interference signal of the two colinear laser frequency combs transmitted through the sample containing the spectral information of the HCHO response is recorded using a fast photodiode (28, 31). The record-short acquisition time of only 0.5 s guarantees constant sample concentration throughout each measurement.

The repetition rates of the dual frequency combs are $f_{rep} = 1$ GHz and $f_{rep} + \delta = 1$ GHz + 19 kHz, respectively. This corresponds to two pulse trains with a sweeping delay between consecutive pulses that causes an interference burst every $1/\delta = 53$ μs, called an interferogram. The optical delay is analogous to the varying pathlength difference in Fourier-transform spectroscopy using a scanning-mirror interferometer, resulting in a down-converted heterodyne signal. After Fourier transformation of the interferogram, the obtained radio frequency spectrum can be up-converted to the UV domain by multiplying the frequency axis with the down-conversion factor $m = f_{rep}/\delta \sim 5 \times 10^4$ (31). With coherent averaging of consecutive interferograms under stable comb conditions, a high signal-to-noise ratio (SNR) is achieved. Simultaneously, through time multiplexing, the required optical power is reduced (25, 31).

An oscilloscope card reads the signal from the photodiode and streams the data of the interferogram chain to storage. A phase-correction algorithm applied post-recording compensates for residual beating frequency drifts due to slow noise components like mechanical resonances. Fast noise components, e.g. from relative intensity noise, are considered to have only minor effects (*33, 37, 38*).

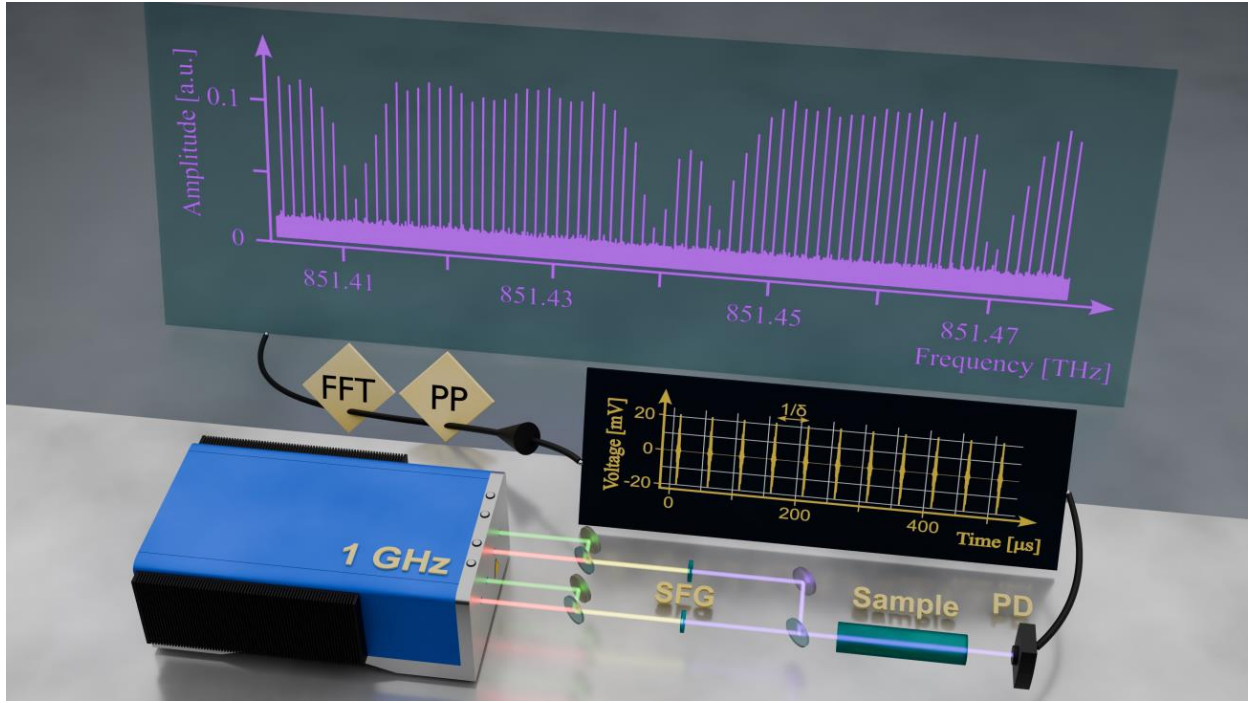

Fig. 1. Experimental scheme of single-cavity dual comb spectroscopy in the ultraviolet. The ultraviolet frequency combs are generated via frequency up-conversion of infrared combs using sum-frequency generation (SFG) between the fundamental laser and the second harmonic. Both UV beams pass through the sample cell containing purified HCHO and generate a time-dependent interference signal on a fast photodiode (PD, see black screen). The time trace is post-processed (PP) using a phase self-correction algorithm. The fast Fourier transformation (FFT) yields transmission spectra with comb-mode resolution (see gray screen).

### *Free-running ultraviolet dual-comb spectrum*

Figure 2A displays the mean spectrum obtained after coherent averaging of 9900 single interferogram bursts. Both frequency combs possess a high mutual coherence with a low relative optical frequency noise, enabling detection of the UV spectrum with a SNR of 255 within 500 ms acquisition time (*33, 39*). The spectrum extends over a bandwidth of ~ 12 THz including more than $10^4$ individually-resolved comb modes and covering more than 600 rovibronic absorption lines of HCHO. The resulting dual-comb quality factor is

$$QF = \text{SNR} \times \frac{\Delta f}{\delta f \cdot \sqrt{T}} = 255 \times \frac{12.4\,\text{THz}}{1\,\text{GHz} \cdot \sqrt{0.5\,\text{s}}} \approx 4.5 \times 10^6 \sqrt{\text{Hz}},$$ with a spectral resolving power of ~$10^6$,

the highest reported value to date (*40*). Figure 2B shows the comb-mode-resolved spectrum with five characteristic HCHO absorption features. The frequency comb modes have transform-limited linewidth and are spaced by 1 GHz, corresponding to the repetition rate of the laser (*37*).

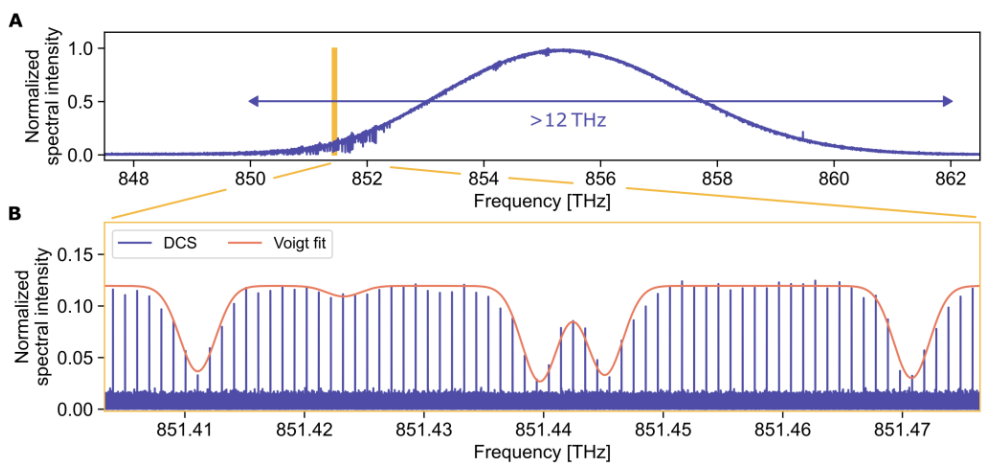

Fig. 2. Broadband near-ultraviolet dual-comb spectrum. (A) The near-ultraviolet, coherently-averaged spectrum is centered at 855.4 THz and covers a full bandwidth of 12.4 THz, resolving the congested absorption features between 850 THz and 852.5 THz and around 855 THz. (B) Comb-mode-resolved absorption spectrum of the phase-corrected signal using a 500 ms-long trace. The absorption dips of HCHO are sampled with a spectral resolution of 1 GHz.

### *Broadband absorption spectroscopy of formaldehyde*

To benchmark our results, Figure 3 compares the UV DCS absorption spectrum to state-of-the-art measurements with the highest reported resolution in this spectral window (*18, 22*). The linewidths of the absorption features measured in this work agree with previous studies and the predicted width assuming Doppler broadening (T = 294 K). We observe a minor relative frequency offset on the order of $10^{-6}$ in the line positions between the DCS measurement and the literature (see Fig. 3B & F). We use a previously-reported transition frequency (848.79 THz, $4^1_0$) from literature as a reference (*41*). Due to the high SNR of 255, DCS permits the detection of previously unresolved weak absorption lines down to $10^{-22}$ cm$^2$/mol, which are observed at frequencies above 851.8 THz (see Fig. 3C-E & G-I, note the different y-scale). The rotational linewidths and positions of these lines are investigated for the first time because of an improvement in SNR and spectral resolution compared to state-of-the-art of more than one order of magnitude.

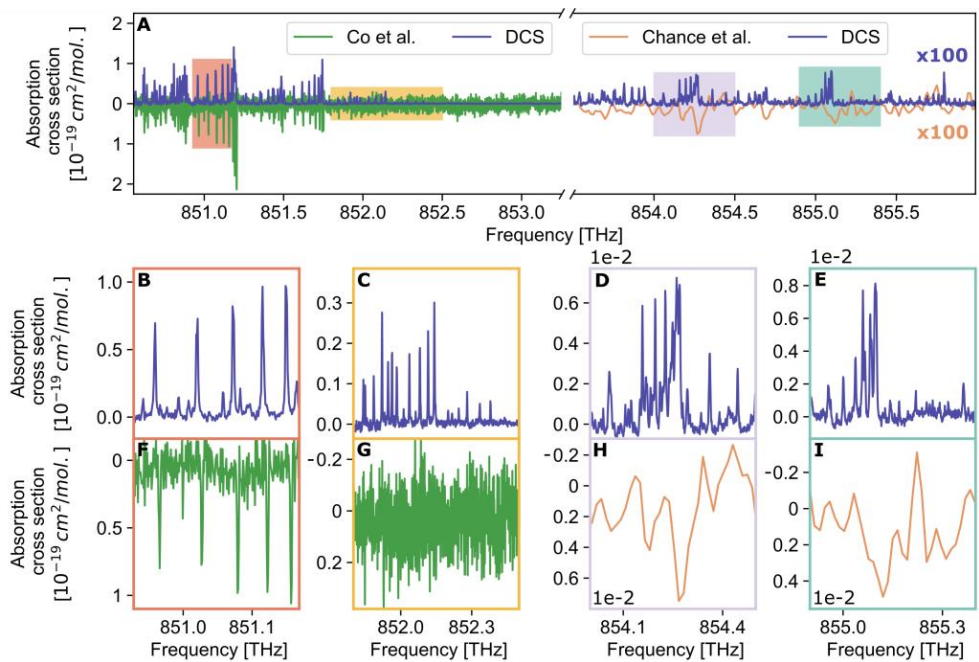

Fig. 3. Comparison of ultraviolet absorption cross section data. (A) DCS absorption cross section spectrum of HCHO (blue, above the abscissa) compared to state-of-the-art measurements (*18*, *22*). The strong absorption features of both traces agree qualitatively but display minor differences in the line positions (see (B) and (F)). The DCS spectrum has an up to 20 times lower noise level (see (C) and (G)). For frequencies above 853 THz, the DCS trace resolves the rovibronic lines for the first time (see (D), (E) and (H), (I)). The mean relative absorption cross section uncertainty amounts to 7.2 %.

### *Absolute absorption cross sections of formaldehyde*

To substantiate the recorded absorption spectrum with theory, we perform simulations of the absorption cross section spectrum of HCHO using PGOPHER (*42*) with the rotational parameters, including centrifugal distortion constants, from Ref. (*41*, *43*). The simulation enables assigning the corresponding transitions, including all relevant quantum numbers. We then carried out a new fit using our data. Knowledge about the line positions and intensities allows for precise testing of rotational constants and investigation of perturbations and higher-order corrections (*44*).

Figure 4 shows the comparison between the simulation using our new constants and the experimental DCS spectrum. The two spectra are in excellent agreement, as evidenced by low residual values (see Fig. 4B). The line positions agree with a standard deviation of 0.37 GHz, significantly below the spectral resolution (see Fig. 4C). The full-width-at-half-maxima of the Voigt line fits are in excellent agreement with the linewidth expected from Doppler broadening (T = 294 K), as the example in Fig. 4D shows.

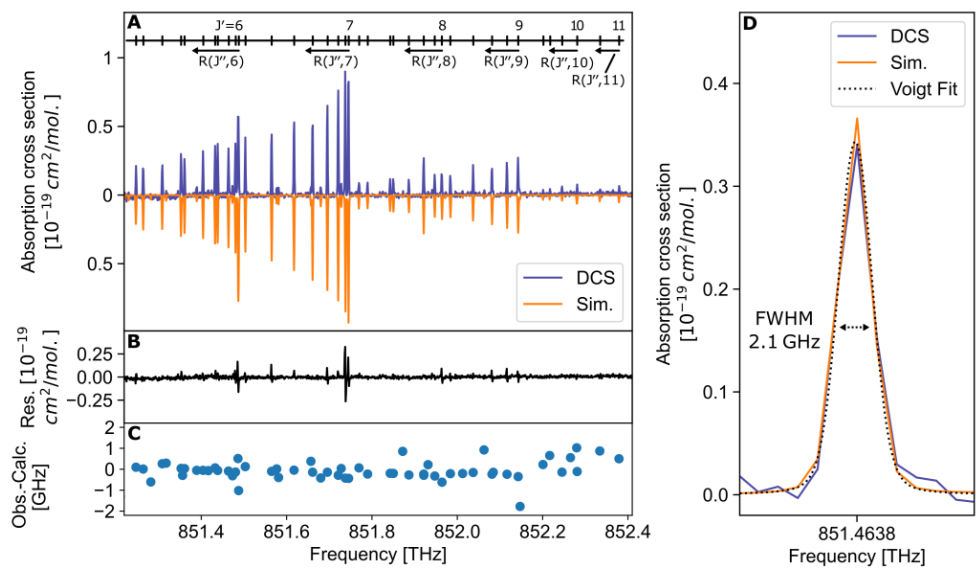

Fig. 4. Comparison between simulation and dual comb spectroscopy data. (A) DCS absorption cross section of HCHO ($4^1_0$-vibronic branch, T=294 K, 310 cm pathlength and 1.07 mbar pressure) in comparison to a simulated spectrum (Sim.) (*42*). (B) The residual values (Res.), i.e., the difference between the DCS and simulation curve, reveal minor differences in the line intensities. (C) The obtained line positions show excellent agreement. (D) A single absorption line centered at 851.404 THz and the corresponding Voigt fit. The full-width-at-half-maximum (FWHM) amounts to 2.1 GHz, in good agreement with the expected width assuming Doppler broadening.

Figure 5 displays the absolute absorption cross section of HCHO between 853.2 THz and 856 THz. We detect a complex structure of rotationally-resolved absorption features, characteristic for ultraviolet absorption spectra where electronic, vibrational, and rotational degrees of freedom are involved, which makes an unambiguous assignment challenging. This absorption spectrum constitutes the first experimental resolution of rotational transitions in HCHO (see Figure 3H). The corresponding simulation and measurement data are provided in the supplementary material. We assign the absorption lines to the weak $3^1_0 4^2_1$ vibronic branch, and the simulation is in good agreement with the measured DCS trace using improved rotational constants A, B, C, and centrifugal distortion terms for this branch, helping to refine quantum mechanical models (see Supplementary) (*45*). These analysis efforts extend the knowledge about the electronic fingerprint and molecular structure of HCHO in the important ultraviolet region further. Figure S3 shows our absolute absorption cross section data compared to the simulation using the currently listed constants (*45*) for visualization of the dramatic improvement.

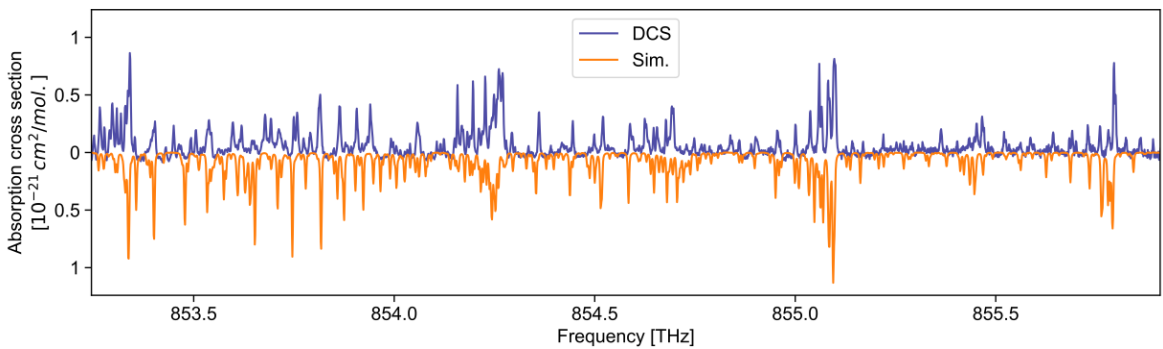

Fig. 5. High-resolution rovibronic absorption spectrum of formaldehyde. Measured absorption cross section spectrum of HCHO (DCS, T=294 K, 310 cm pathlength, and 90 mbar pressure) including rotational states that have not been experimentally observed to date. The measured absorption lines do not obey periodic branching like in the $4^1_0$ band and have an absorption cross section on the order of $10^{-22}$ cm$^2$/mol. The simulation (Sim.) is performed with our improved simulation parameters compared to the literature, fitting to the experiment (see Supplementary) (*42*, *45*).

## *3. Discussion*

In conclusion, this work reports the investigation of the photochemically relevant ultraviolet absorption cross section spectrum of HCHO around 855 THz (350.6 nm wavelength) at a sensitivity and resolution level that allows for the refinement of molecular simulation parameters. We employ a free-running, single-cavity dual-comb laser system and uncover hundreds of previously undetected rovibrational formaldehyde transitions. The rotationally-resolved absolute absorption cross section spectra yield a precise electronic fingerprint of HCHO. The absorption line positions were measured with a relative accuracy better than $10^{-6}$, and for frequencies beyond 853 THz, we demonstrate an improvement in spectral resolution of more than one order of magnitude over state-of-the-art measurements.

Atmospheric processes include dissociation channels for frequencies above 833 THz ($\lambda < 360$ nm wavelength) via excited electronic states and photochemical or photophysical oxidation, where knowledge about the absolute absorption cross section is inevitable (*46–48*). Our work adds significantly to the knowledge of the formaldehyde behavior in this photochemically-active spectral region that plays an important part in crucial tropospheric pollution monitoring projects like TEMPO, TROPOMI and GEMS (*14*, *15*, *17*). With our compact and robust setup, precision spectroscopy is feasible, and mobile field measurements for environmental sensing in the ultraviolet spectral region come into reach. We estimate the HCHO detection limit of the presented system to 3 ppb for an interaction pathlength of 5 km, typical for atmospheric sensing. Given the standard tropospheric abundance of formaldehyde between 1-4 ppb, the presented system is ready for field operation when combined with a sampling path setup similar to Eber et al. (*49*). Our new system does not require additional optical stabilization schemes and can be operated without extensive electronic supplies. In terms of simplicity, only Xu et al. (*25*) could report on a simplified, EOM-based UV dual-comb spectrometer that achieves comb-resolved spectra. We realize a spectral bandwidth of more than 12 THz, spanning across numerous absorption features of HCHO. Investigating HCHO directly in the atmosphere would yield valuable information about its complex photochemistry and pre-dissociation, especially in areas where precarious concentrations are found, e.g. industrial areas and biomass burnings (*9*, *48*, *50*).

Enhancing the spectral bandwidth to frequencies above 884 THz (339 nm wavelength), where high absorption cross sections of up to $4\times10^{-19}$ cm$^2$/mol. are observed, paves the way towards ultra-sensitive detection of formaldehyde even at the ppt-level. This becomes achievable with direct spectral broadening in the UV (28). Furthermore, literature values of the absorption cross section in the extended photochemically active spectral region still differ significantly up to now (46, 51). An alternative detection method with high sensitivity and unprecedented resolution, as demonstrated here, will help resolve this spectroscopic conundrum, which is decisively relevant for environmental monitoring at these very low levels.

The differences in line intensities between measurement and simulation can be explained by perturbations such as Coriolis-type interactions between different vibrational modes. Through the calibration of the absorption line intensity, our absolute absorption cross section data warrants a precise analysis that helps toward the understanding of processes such as pre-dissociation (50).

In comparison to state-of-the-art methodology in ultra-resolution spectroscopy in the UV, which includes synchrotron laboratories, tunable frequency up-converted laser spectroscopy, and Fourier-transform spectroscopy, our method provides a compact, cost-effective, and rigid measurement concept with record short acquisition times and unparalleled spectral resolution. This renders UV dual-comb spectroscopy suitable to decisively advance environmental sensing applications (49), (laboratory) astrophysics (52), and medical diagnosis (53, 54).

## Acknowledgments

The authors gratefully acknowledge support from NAWI Graz. We thank Monika Mayer, Harald Rieder and Jonathan Tennyson for valuable discussions.

**Funding**

NAWI Graz Core Facility Infrastructure Program

European Research Council (ERC) under the European Union's Horizon 2020 research and innovation programme grant No 947288

Austrian Science Fund FWF START programme grant No Y1254

**Competing interests:** The authors declare that they have no competing interests.

**Data and materials availability:** All data needed to evaluate the conclusions in the paper are present in the paper and/or the Supplementary Materials. The data that support the findings of this study are available from the corresponding authors upon request. The code used to generate the figures is available from the corresponding author upon request.

# Supplementary Materials

**Materials and Methods**

Dual-comb system and harmonic generation

The commercially available laser source in this work is based on a Yb:CALGO crystal-based, single-cavity dual oscillator with 285 THz center frequency (1052 nm wavelength). The two optical frequency combs from a single laser cavity are spatially multiplexed using a Brewster-angled biprism. The laser system operates in free-running mode without active stabilization of the comb parameters. The repetition rates are measured to $f_{rep,1} = 1$ GHz and $f_{rep,2} = 1$ GHz + 19 kHz.

For each of the two infrared combs, a waveplate and a polarizing beam splitter send a variable ratio of the fundamental radiation to a nonlinear crystal for second harmonic generation. The harmonic radiation is spatially superposed with the residual infrared beam using a dichroic mirror. Then, both beams are focused by an off-axis parabolic mirror into a second nonlinear crystal for sum-frequency generation. The UV light is re-collimated by a lens and separated from the other beams via dichroic mirrors. The resulting spectrum is centered at 855.4 THz frequency (350.5 nm wavelength) and the pulses have approximately 5 pW pulse energy. The two ultraviolet beams are spatially superposed at a plate beamsplitter and the interferogram is detected using a fast photodiode. Besides the sample cell, the whole beam path is in air. We define the baseline for absorption measurements using a baseline fitting routine. Assuming linearity of the Beer-Lambert law, we calculate the absorption cross section values using the HCHO pressure, the length of the absorption cell, and laboratory temperature (*31*). We use a gas-type independent capacitance pressure gauge and a purified, liquid reservoir of monomeric HCHO (see section preparation of monomeric formaldehyde).

Sample preparation

Monomeric formaldehyde is prepared in a two-step purification process. The resulting liquid is transferred to a stainless-steel reservoir and connected to the vacuum setup. The whole vacuum setup is heated and evacuated overnight to warrant low contamination, e.g. from water. The sample cell is evacuated to the $10^{-2}$ mbar range. Then, we let gas-phase HCHO expand into vacuum until the desired pressure is reached, which is monitored by a calibrated, gas-type-independent capacitance pressure gauge. The pressure gauge is calibrated, i.e. zero-adjusted, at a background pressure of less than $10^{-6}$ mbar and has an accuracy of 0.2 % of the measured value. The optical path through the sample cell has a length of 62 cm with windows oriented at Brewster's angle. For the results presented, a multi-pass geometry with 5 passes was used, which yields a total interaction length of 310 cm. The pressure is monitored during measurements for precise determination of the HCHO concentration.

Preparation of monomeric formaldehyde

HCHO is an irritant chemical, which is toxic by skin contact and inhalation, and is a known carcinogen. Therefore, its preparation and disposal have to be carried out in a well-ventilated fume hood wearing personal protective equipment (lab coat, safety glasses, nitrile protective gloves).

This setup was adapted from previous works (*21-23*), but instead of a specialized apparatus commercially-available glass equipment employed in synthetic chemistry laboratories is used. The cracking and distillation process is performed in a single apparatus.

The experimental setup for the cracking of paraformaldehyde involves a glass apparatus composed of a Schlenk flask (see Fig. S1, left) connected by a glass bridge to a cooling trap (see Fig. S1, middle), which is again connected via a glass bridge to a receiving Schlenk flask (see Fig. S1, right). The glass apparatus is connected to a Schlenk-type argon/vacuum manifold. Before performing the cracking process, about 5 g paraformaldehyde (extra pure, Merck, K12297505) is dried in a NS29 Schlenk flask by heating in an oil bath at 80 °C under vacuum (~1 mbar) for 16 h followed by 2 h at 100 °C.

The glass apparatus depicted in Figure S1 is evacuated to < 0.5 mbar and dried thoroughly using a heat gun. After ventilation with argon the Schlenk flask containing the dried paraformaldehyde is connected (see Fig. S1, left) and the glass apparatus again evacuated and heated thoroughly. After the apparatus cools to room temperature under vacuum, the first cold trap is cooled in a Dewar with liquid nitrogen. The Schlenk flask containing the paraformaldehyde is immersed in an oil bath (see Fig. S1), which is gradually heated to 150 °C. This process should be performed carefully in order to minimize paraformaldehyde particles, which might start "jumping" in the course of the depolymerization process due to gas evolution. During the entire cracking process, the oil pump vacuum holds the receiving Schlenk flask at ~ 0.6 mbar. The monomeric formaldehyde, as well as residual water, collects in the upper region of the intermediary cold trap as a white solid.

After a sufficient amount of paraformaldehyde has depolymerized, the oil bath is removed and the cryogenic distillation of the monomeric formaldehyde is pursued. The receiving Schlenk flask is cooled in a Dewar filled with liquid nitrogen (-196 °C) and the liquid nitrogen Dewar of the intermediary cooling trap is subsequently replaced with a Dewar filled with dry ice/acetone (-78 °C).

The connection to the vacuum pump is closed. The solidified formaldehyde in the cooling trap melts as it warms to the temperature of the dry ice/acetone bath and distills into the receiving Schlenk flask. In order to maintain vacuum in the apparatus, every 5-10 min the connection to the vacuum pump is opened shortly. After a sufficient amount of formaldehyde has been distilled, it is transferred via a cannula into a specially-designed stainless-steel reservoir possessing a septum and an integrated hose-adapter.

Preparing for the transfer, the stainless-steel vessel is connected to a second Schlenk line via the integrated hose adapter and rendered inert by heating under vacuum with a heat gun. Then, continuous argon flow is applied. A cannula, which has been freshly taken from a drying oven at 120 °C, is plunged through the septum, and the lower collection compartment is cooled to -78 °C in an dry ice/acetone bath. Argon flow is maintained through the cannula up until the cannula transfer of the liquid.

In parallel, the glass apparatus is flushed with argon through the connection to the receiving Schlenk flask. While maintaining constant argon flow, the receiving flask is disconnected from the apparatus, closed with a rubber septum and immediately immersed into a Dewar filled with dry ice/acetone (-78 °C). At this temperature, the distilled formaldehyde forms a colorless liquid. In order to avoid the release of formaldehyde vapors from the unused part of the cracking and distillation glass apparatus, a dummy flask is used to close off the glass apparatus again.

The liquid from the receiving Schlenk flask is transferred to the stainless-steel vessel by plunging the other end of the cannula through the septum of the Schlenk flask. By applying overpressure of argon in the Schlenk flask and providing slight pulses of vacuum to the stainless-steel vessel, the transfer succeeds within seconds. The stainless-steel vessel is disconnected from the Schlenk line and transported in a dry ice/acetone Dewar to the laser apparatus. It remains in this Dewar during the entire period of the measurements.

Disposal of formaldehyde
Any remaining formaldehyde in the receiving flask and the stainless-steel vessel is carefully quenched by the addition of distilled water under argon at -78 °C. A bubbler containing 0.1 M NaOH is attached before the flasks are left to warm up to room temperature overnight. The used glassware pieces are thoroughly cleaned from repolymerized paraformaldehyde with 0.1 M NaOH, an NaOH/iPrOH base bath (24 h), and a 0.1 M HCl bath (24 h).

Free-running dual-comb characterization
We characterize our dual-comb spectrometer by investigating the scaling of the signal-to-noise-ratio (SNR) versus measurement time and average power on the photodiode. Figure S2A shows that the SNR scales with the square root of measurement time. The SNR increases for higher average powers on the detector with a linear dependence.

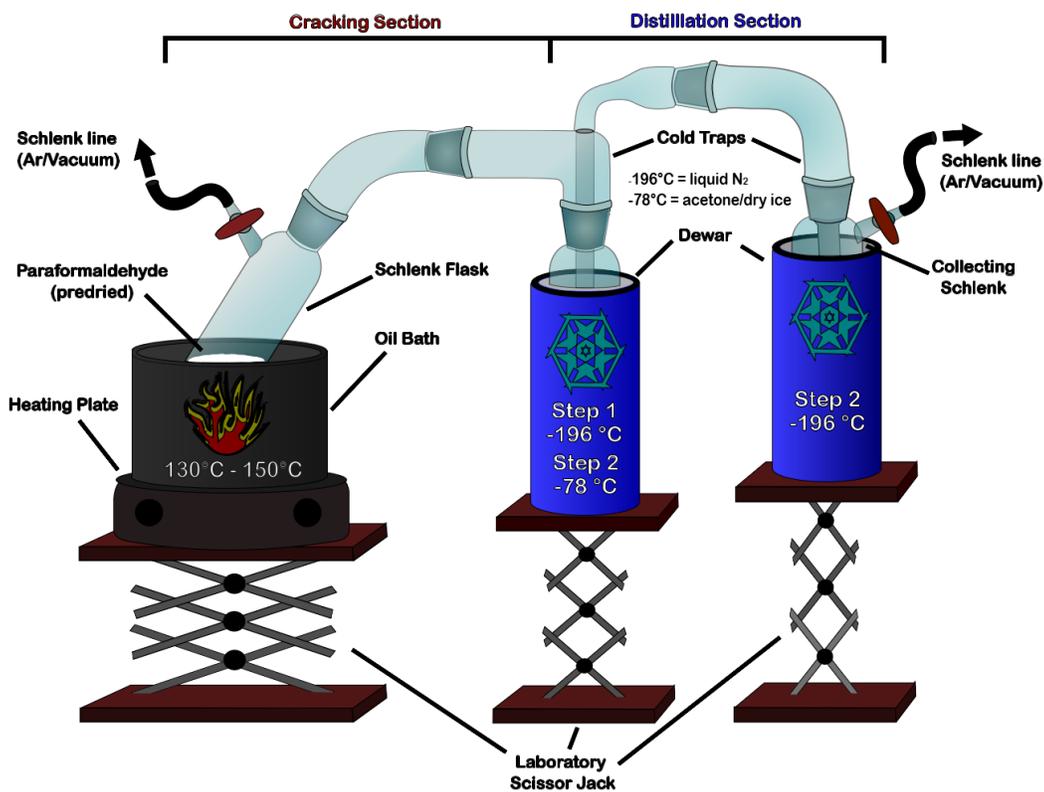

Fig. S1. Experimental apparatus for the preparation of monomeric formaldehyde. Formaldehyde is collected in the cooling trap in the middle, which is cooled to -196°C using liquid nitrogen during the cracking process (Step 1). For the purification process (Step 2) the formaldehyde from the cooling trap is distilled into the receiving Schlenk flask, which is cooled in a Dewar to -196°C.

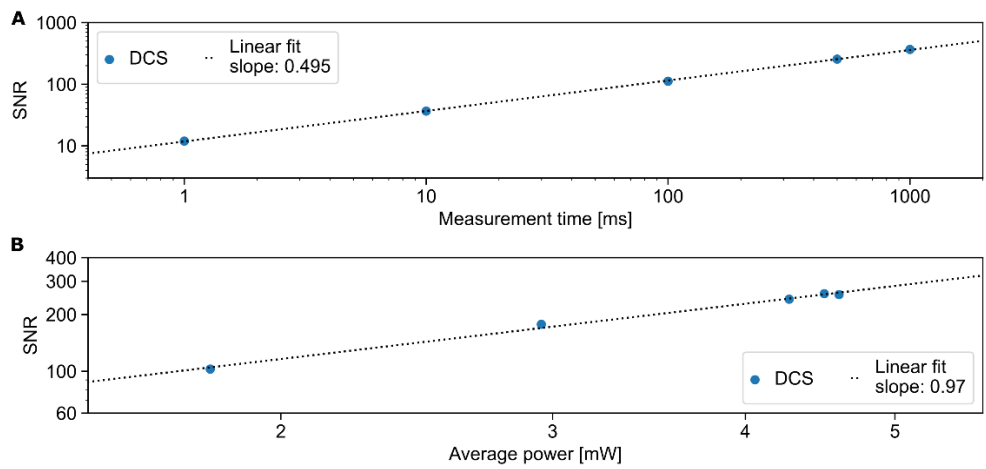

Fig. S2. Signal-to-noise-ratio (SNR) of free-running dual comb spectroscopy. (A) The SNR is plotted versus measurement time and follows a square-root scaling as expected. (B) The SNR increases for higher average powers at the detector.

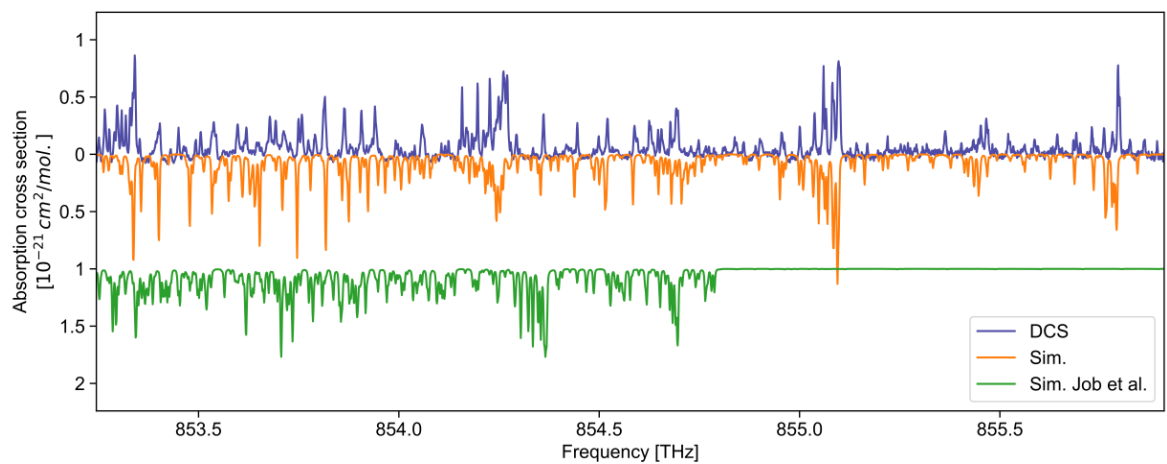

Fig. S3. High-resolution rovibronic absorption spectrum of formaldehyde. Measured absorption cross section spectrum of HCHO (DCS, T=294 K, 310 cm pathlength, and 90 mbar pressure) including rotational states that have not been experimentally observed to date. The simulation (Sim.) is performed with updated constants and the reported simulation parameters from Job et al. (*45*).

Table T1.

Simulation constants of the $4^1_0$ branch (see Fig. 4) and the $3^1_0 4^2_1$ branch (see Fig. 5).

| Vibronic branch | $4^1_0$ | $3^1_0 4^2_1$ |
|---|---|---|
| Origin [MHz] | 848789501.50652 | 853410070.14698 |
| A [MHz] | 262375.79522915 | 259446.301702701 |
| B [MHz] | 33719.6177988498 | 44660.7250457385 |
| C [MHz] | 30325.4459294014 | 15855.1426714503 |
| DK [MHz] | 15.0101921239566 | -2.91521639571955 |
| DJK [MHz] | 1.96896891997144 | -10.5803794984655 |
| DJ [MHz] | 0.0650972482888758 | -16.5593831059873 |
| ΔK [MHz] | -3.2941130518192 | 316.272078265534 |
| ΔJ [MHz] | 0.0666183905677189 | 46.4430426652569 |